\documentclass[10pt,a4paper]{article}

\usepackage[T1]{fontenc}
\usepackage[utf8]{inputenc}
\usepackage{times}
\usepackage[english]{babel}
\usepackage{url}
\usepackage{amsmath,amsthm,amssymb,color,listings,algorithm,algpseudocode}
\usepackage{a4wide}
\usepackage{multicol}
\usepackage{listings}
\usepackage{graphicx}
\usepackage{caption}
\usepackage{subcaption}
\usepackage{float}
\usepackage[
  pdfencoding=auto,
  colorlinks=true,
  linkcolor=black,
  urlcolor=black,
  citecolor=black]{hyperref}
\usepackage{tikz}
\tikzset{every node/.style={fill=white}}
\usepackage{physics}
\usepackage{pdflscape}
\usepackage{array,multirow}
\usepackage{relsize}
\usepackage{pxfonts}
\usepackage{pdftexcmds}
\usepackage{pgfplots}
\usepackage{empheq}
\pgfplotsset{compat=1.15}
\usepackage{enumitem}
\setlist{nosep}
\usepackage{diagbox}
\usepackage{mathtools}
\usepackage{authblk}
\newcommand\blfootnote[1]{%
  \begingroup
  \renewcommand\thefootnote{}\footnote{#1}%
  \addtocounter{footnote}{-1}%
  \endgroup
}

\newtheorem{definition}{Definition}

\newtheorem{example}{Example}
\newtheorem{remark}{Remark}

\providecommand{\keywords}[1]{\vspace{.5cm}
\noindent\textbf{\textit{Keywords ---}} #1}

\newcommand\add[1]{#1}
\newcommand\modif[1]{#1}
\newcommand\forMinorRevision[1]{#1}

\algnewcommand\algorithmicforeach{\textbf{for each}}
\algdef{S}[FOR]{ForEach}[1]{\algorithmicforeach\ #1\ \algorithmicdo}
\algnewcommand{\IIf}[1]{\State\algorithmicif\ #1\ \algorithmicthen}
\algnewcommand{\EndIIf}{\unskip\ \algorithmicend\ \algorithmicif}
\algnewcommand{\IFor}[2]{\State\algorithmicforeach\ #1\ \algorithmicdo\ #2\ \algorithmicend\ \algorithmicfor}

\def\sec{Sect.}
\def\fig{Fig.}
\def\tab{Tab.}
\def\bigO{O}
\newcommand\eq[1]{Eq.~(#1)}

\title{Contextuality degree of quadrics in multi-qubit symplectic polar spaces}

\author[1]{Henri de Boutray}
\author[2,3,4]{Frédéric Holweck}
\author[2,5]{Alain Giorgetti}
\author[2,5]{\\Pierre-Alain Masson}
\author[6]{Metod Saniga}
\affil[1]{ColibrITD, La Défense, Paris, France}
\affil[2]{Univ. Bourgogne Franche-Comté (UBFC)}
\affil[3]{Laboratoire Interdisciplinaire Carnot de Bourgogne (ICB, UMR 6303),
  Belfort, France}
\affil[4]{Department of Mathematics and Statistics, Auburn University, Auburn, 
  AL, USA}
\affil[5]{Institut FEMTO-ST (UMR 6174 - CNRS/UBFC/UFC/ENSMM/UTBM), Besançon, 
  France}
\affil[6]{Astronomical Institute, Slovak Academy of Sciences, SK-05960 
  Tatransk\'a Lomnica, Slovak Republic}
\date{}

\begin{document}

\maketitle

\begin{abstract}
Quantum contextuality takes an important place amongst the concepts of quantum
computing that bring an advantage over its classical counterpart. For a large
class of contextuality proofs, \textit{aka.} observable-based proofs of the
Kochen-Specker Theorem, we formulate the contextuality property as the absence
of solutions to a linear system and define for a contextual configuration its
degree of contextuality. Then we explain why subgeometries of binary symplectic
polar spaces are candidates for contextuality proofs. We report the results of a
software that generates these subgeometries, decides their contextuality and 
computes their contextuality degree for some small symplectic polar spaces. We
show that quadrics in the symplectic polar space $W_n$ are contextual for
$n=3,4,5$. The proofs we consider involve more contexts and observables than the
smallest known proofs. This intermediate size property of those proofs is
interesting for experimental tests, but could also be interesting in quantum
game theory.

\blfootnote{\hspace{-.55cm} Corresponding author: frederic.holweck[at]utbm.fr\\
This paper was published as \cite{dHG+22}. This is the up-to-date version.}
\end{abstract}

\section{Introduction}
\label{sec:introduction}

In quantum information theory, many paradoxes of the early years of quantum
physics, like superposition or entanglement, have turned on to be considered as
resources for the development of quantum technologies when they exhibit
non-classical behavior. One of these resources is quantum contextuality. The
no-go theorems proved by Kochen-Specker~\cite{KS67} and Bell~\cite{Bel66} are
fundamental results that establish that no Non-Contextual Hidden Variables
(NCHV) model can reproduce the outcomes of quantum mechanics. First demonstrated
as a mathematical result, quantum contextuality has since been tested
experimentally~\cite{ARBC09,KZG+09} and very recently checked on an online
quantum computer~\cite{DRLB20,Hol21}. The importance of contextuality in quantum
computation has been shown in several articles~\cite{HWVE14,Rau13,ORBR17}.

The original proof of the Kochen-Specker Theorem was based on the
impossibility to color bases of rays in a three-dimensional space according to
some constraints imposed by the law of quantum physics. This proof involves
$117$ rays. Many other proofs intending to simplify the initial argument have
also been proposed~\cite{CEG96,WA11,Pla12}.

In the '90s David Mermin~\cite{Mer93} and Asher Peres~\cite{Per90} introduced
a different kind of argument to prove quantum contextuality. Their
observable-based approach is the one that we consider in this paper. We restate
their argument, also known as the ``Mermin-Peres magic square'', as an example
of contextual geometry, in \sec~\ref{sec:geometries}.

The Mermin-Peres magic square and the Mermin pentagram, a three-qubit
observable-based proof of the Kochen-Specker Theorem, have been investigated in
the past 15 years from the perspective of finite geometry. In~\cite{HS17} it
was proven by geometric arguments that these two proofs are the ``smallest''
ones in terms of number of observables and contexts. In~\cite{SPPH07} it was
shown that the 10 possible Mermin-Peres magic squares were hyperplanes of a
point-line geometry known as the doily (see details in 
\sec~\ref{sec:symplectic-space} \forMinorRevision{or in the introduction 
of~\cite{MSG+22}}) and in~\cite{PSH13} the number of different Mermin pentagrams
was obtained and explained later in~\cite{LS17}. Mermin-Peres magic squares have
also been considered from the perspectives of graph theory and binary constraint
system games~\cite{Ark12,CM13}.

Contextuality was formalized using sheaf theory~\cite{AB11}. This formalism
brought many methods to reason on contextuality, in particular Abramsky and
Carù~\cite{AC16} used this framework to study some proofs of contextuality
called \emph{All-versus-Nothing} (abbreviated \emph{AvN}) proofs, and to
construct these proofs using a computer program. We do not intend to improve on
these results but rather to enumerate some of these AvN proofs as special cases
of finite geometries. Other authors have used a correspondence between
contextuality and a linear problem initially introduced in~\cite{AB11} to
provide a global AvN enumeration (see, \textit{e.g.}~\cite{ABCP17}) or links
between contextuality and logical paradoxes, as in~\cite{Bel64}.

In this paper we address automatic checking of observable-based proofs of the
Kochen-Specker Theorem with higher numbers of contexts and observables. In
particular we check that the symplectic polar spaces $W_n$ of rank $n$ and order
$2$ are contextual for $n=2,3,4$, when seen as point-line geometries encoding
the commutation relations in the $n$-qubit Pauli group. We also check that all
hyperbolic and elliptic quadrics, which are subgeometries of $W_n$ defined by
quadratic equations, are contextual, again for $n=2,3,4$. Despite the fact that
elliptic and hyperbolic quadrics and their connection with multiple qubits Pauli
observables have already been studied in the quantum information
literature~\cite{SGL+10,LHS17}, the contextuality property of those
configurations has not been established before. Because those configurations
involve a lot of observables and contexts (for instance 135 observables and 1575
contexts for $n=4$), we use a computer software to check their contextuality and
to compute the degree of contextuality of some of those configurations. Looking
at observable-based proofs of contextuality with large numbers of observables
and contexts can be interesting to build macroscopic state-independent
inequalities that violate non-contextual hidden variables
inequalities~\cite{Cab08,Hol21}. Another motivation comes from quantum game
theory, as more sophisticated proofs  could lead to more complex game scenarios
than for instance the Magic square game~\cite{BBT05}.

In \sec~\ref{sec:geometries} we recall the perspective of finite geometry on
observable-based proofs of the Kochen-Specker Theorem. In \sec~\ref{sec:degree}
we recall how these proofs translate to the resolution of a linear system over
the two-element field $\mathbb{F}_2$ and define the notion of degree of
contextuality for an observable-based proof of the Kochen-Specker Theorem. In
\sec~\ref{sec:symplectic-space} we recall how the symplectic polar space $W_n$
of rank $n$ and order $2$ encodes the commutation relations in the $n$-qubit
Pauli group, and we explain how contextual configurations of observables live in
$W_n$ as subgeometries. In \sec~\ref{sec:contextual-subspaces} we precisely
define the subgeometries characterized by quadratic equations, \add{we explain
how our program generates these geometries and checks their contextuality,} and
we provide \modif{its contextuality results}. We show in particular that all
quadrics of $W_n$ define contextual configurations for $n=3,4,5$ \add{and we
compute their degree of contextuality for $n=3$}. In
\sec~\ref{sec:quadric_geometry} we look at \modif{this new result for $n = 3$}
from a geometric perspective \add{and show how this knowledge is useful to
characterize contextual inequalities, with the perspective to test contextuality
on a quantum device}. \add{\sec~\ref{sub:cayley} presents another application of
our software.} \sec~\ref{sec:conclusion} is dedicated to concluding remarks. The
code mentioned in this article is publicly available at
\url{https://quantcert.github.io/Magma-contextuality}.

\section{Contextual geometries}
\label{sec:geometries}

We first propose a precise definition of the notion of contextual geometry,
underlying the previous work on the geometrical perspective on observable-based
proofs of the Kochen-Specker Theorem. Our definition may not be as general as
possible, but it is sufficient for the present work. We illustrate it with the
well-known example of the Mermin-Peres magic square. Then we reformulate the
contextuality property as inconsistency of a linear system with coefficients in
the field of modulo-2 arithmetic.

A \emph{quantum geometry} is a pair $(O,C)$ where $O$ is a finite set of
\emph{observables} \modif{(finite-dimensional Hermitian operators)} and $C$ is a
finite set of subsets of $O$, called \emph{contexts}, such that 
\begin{enumerate}[label=\textbf{O.\arabic*}]
  \item each observable $M \in O$ satisfies $M^2=Id$ (so, its eigenvalues are in 
    $\{-1,1\}$);
  \label{enum:binary}
  \item any two observables $M$ and $N$ in the same context commute, {\it i.e.},
    $MN=NM$;
  \label{enum:op-commut}
  \item the product of all observables in each context is either $Id$ or $-Id$.
  \label{enum:context-prod}
\end{enumerate}
The elements of $O$ and $C$ are the \emph{points} and \emph{lines} of the
geometry. Let the \emph{context valuation} associated to $(O,C)$ be the function
$e : C \rightarrow \{-1,1\}$ defined by $e(c) = 1$ if the product of all the
observables in the context $c$ is $Id$, and $-1$ if it is $-Id$.

A \emph{contextual geometry} is a quantum geometry such that there is no
\emph{(observable) valuation} $f : O \rightarrow \{-1,1\}$ such that
\begin{equation}
\label{eq:contextual-geom}
\forall c \in C, \ \prod_{M \in c} f(M) = e(c). 
\end{equation}
\noindent This statement is equivalent to the assertion that there is no NCHV
model that could explain the result predicted by quantum mechanics for this
geometry. If the geometry is not a contextual geometry, it is said to be
\emph{non-contextual}.

\subsection{Example: Mermin-Peres magic square}

We denote by $X$, $Y$ and $Z$ the usual Pauli matrices, \textit{i.e.}
\begin{equation}
 X=\begin{pmatrix}
    0 & 1\\
    1 & 0
   \end{pmatrix},Y=\begin{pmatrix}
   0 & -i\\
   i & 0
   \end{pmatrix} \text{ and } Z=\begin{pmatrix}
   1 & 0\\
   0 &-1
\end{pmatrix},
\end{equation}
and by $I$ the $2\times 2$ identity matrix.

\begin{figure}[!ht]
\begin{center}
\begin{tikzpicture}[scale=1.4, 
]
\node (1) at (0,0) {$X \otimes Y$};
\node (2) at (1,0) {$Y \otimes X$};
\node (3) at (2,0) {$Z \otimes Z$};
\node (4) at (0,1) {$I \otimes Y$};
\node (5) at (1,1) {$Y \otimes I$};
\node (6) at (2,1) {$Y \otimes Y$};
\node (7) at (0,2) {$X \otimes I$};
\node (8) at (1,2) {$I \otimes X$};
\node (9) at (2,2) {$X \otimes X$};
\draw (1) -- (4) -- (7) ;
\draw (2) -- (5) -- (8) ;
\draw [double distance = 2pt] (3) -- (6) -- (9) ;
\draw (1) -- (2) -- (3) ;
\draw (4) -- (5) -- (6) ;
\draw (7) -- (8) -- (9) ;
\end{tikzpicture}
\end{center}
\caption{The Mermin-Peres magic square: There is no NCHV model, i.e. no classical 
  function that can reproduce the outcomes predicted by quantum physics unless
  the function is context dependent.}
\label{fig:mermin_square}
\end{figure}

Let us present the Mermin-Peres magic square as a quantum geometry and restate
the argument of Mermin and Peres for its contextuality. The original
configuration of nine two-qubit observables proposed by
Mermin~\cite[\sec~V]{Mer93} is showcased in \fig~\ref{fig:mermin_square}.
It is the quantum geometry $(O,C)$ where $O = \{ X \otimes I, I \otimes X, X
\otimes X, I \otimes Y, Y \otimes I, Y \otimes Y, X \otimes Y, Y \otimes X, Z
\otimes Z \}$ and $C = \{c_1, c_2, c_3, c_4, c_5, c_6\},$ with $c_1 = \{X
\otimes I, I \otimes X, X \otimes X \}$, $c_2 = \{I \otimes Y, Y \otimes I, Y
\otimes Y\}$,  $c_3 = \{X \otimes Y, Y \otimes X, Z \otimes Z\}$,  $c_4 = \{X
\otimes I, I \otimes Y, X \otimes Y\}$, $c_5 = \{I \otimes X, Y \otimes I, Y
\otimes X\}$ and $c_6 = \{X \otimes X, Y \otimes Y, Z \otimes Z\}$. These six
sets of observables are contexts, since they contain mutually commuting
observables whose product is $\pm Id$. In this example the context valuation $e$
is such that $e(c_1) = \ldots = e(c_5) = 1$ and $e(c_6)=-1$. In
\fig~\ref{fig:mermin_square} the $5$ \emph{positive} contexts $c_1$ to $c_5$
whose product of observables is $+Id$ are depicted as simple lines, whereas the
\emph{negative} context $c_6$ whose product of observables is $-Id$ is depicted
as a double line.

Since the eigenvalues of each observable are $\pm 1$ and the measurements on
each context are compatible (because the observables are mutually commuting) the
product of the observed eigenvalues should be equal to the eigenvalue of the
product of observables. In other words when a context is positive (resp.
negative), \textit{i.e.}, when the product of its observables is $+Id$ (resp.
$-Id$), then quantum mechanics says that the product of the observed
measurements should be $+1$ (resp. $-1$).

Now, a simple argument shows that there is no non-contextual, \textit{i.e.} not
context-dependent, deterministic function $f$ that can assign an outcome $\pm 1$
to each observable and satisfy at the same time the $6$ constraints: If one
multiplies all together the outcomes of the $6$ contexts given by a
non-contextual deterministic function, the result will be $+1$ because each node
will show up twice in the product. However, because there is only one negative
context in the Meres-Peres magic square, this product should be $-1$ if all
constraints are satisfied.

The observable valuation $f$ in the contextuality
property~(\ref{eq:contextual-geom}) formalizes the NCHV hypothesis. If there
were to be a deterministic non-contextual theory explaining the outcomes of
quantum physics, there would be some processes in Nature, hidden from us, that
would allow us to calculate these outcomes. Those hidden processes are generally
called hidden variables and here $f$ is a non-contextual function  which could
depend on those hidden variables: for a set of hidden variables $\lambda$,
$f(M)$ is a shortcut for $f(\lambda,M)$.

\section{Degree of contextuality}
\label{sec:degree}

Let $\mathbb{F}_2 = (\{0,1\},+,\times)$ be  the two-elements field of
modulo-2 arithmetic. Let $(O,C)$ be a quantum geometry with $p = |O|$
observables/points $\{M_1,\ldots,M_p\}$ and $l = |C|$ contexts/lines
$\{c_1,\ldots,c_l\}$.  Its \emph{incidence matrix} $A \in \mathbb{F}_2^{l \times
p}$ is defined by $A_{i,j} = 1$ if the $i$-th context $c_i$ contains the $j$-th
observable $M_j$. Otherwise, $A_{i,j} = 0$. Its \emph{valuation vector} $E \in
\mathbb{F}_2^{l}$ is defined by $E_i = 0$ if $e(c_i) = 1$ and $E_i = 1$ if
$e(c_i) = -1$, where $e$ is the context valuation of $(O,C)$.

With these notations, the quantum geometry $(O,C)$ is contextual iff the linear
system
\begin{equation}
\label{eq:all-pc}
A x = E
\end{equation}
has no solution in $\mathbb{F}_2^{p}$. The matrix $A$ being of size $l\times p$
with $l\leq p$, \eq{\ref{eq:all-pc}} can be efficiently solved with a complexity
$\bigO(p^3)$, \textit{e.g.} by Gaussian elimination.

Note that $A$ is built from the incidence structure of the quantum
geometry $(O,C)$ while the vector $E$ comes from the signs of the contexts. In
other words the left-hand side of \eq{\ref{eq:all-pc}} only depends on the
geometric structure -- that will be revisited in the next section -- while the
right-hand side depends on the outcomes predicted by quantum physics.

In this setting one can measure how much contextual a given quantum geometry is.
Let us denote by $\text{Im}(A)$ the image of the matrix $A$ as a linear map
$A:\mathbb{F}_2^{p}\to \mathbb{F}_2^l$. Then, if $(O,C)$ is contextual,
necessarily $E \notin \text{Im}(A)$. One can define a natural measure of the
degree of contextuality of $(O,C)$ as follows:

\begin{definition}[Contextuality degree]
Let $(O,C)$ be a contextual geometry with valuation vector $E\in \mathbb{F}_2^l$.
Let us denote by $d_H$ the Hamming distance on the vector space $\mathbb{F}_2^l$.
Then one defines the degree $d$ of contextuality of $(O,C)$ by
 \begin{equation}
  d=d_H(E,\text{Im}(A)).
 \end{equation}
\end{definition}

\begin{example}
If one considers the Mermin-Peres magic square then the degree of contextuality
of the configuration is $1$. Indeed in this case $E$ is the column vector
$(0~0~0~0~0~1)^T$ and the column vector $E'=(0~0~0~0~0~0)^T$ is in Im$(A)$. So
the degree is at most $1$. But because the configuration is contextual, one knows
the degree is at least $1$. One concludes that the degree is exactly $1$.
\end{example}

The notion of degree of contextuality measures in some sense how far is a given
configuration to be satisfiable by an NCHV model. The Hamming distance tells us
what is the minimal number of constraints on the contexts that one should change
to make the configuration valid by a deterministic function. In other words
$l-d$ measures the maximum number of constraints of the contextual geometry that
can be satisfied by an NCHV model.

The degree of contextuality has also a concrete application as it can be used to
calculate the upper bound for contextual inequalities. Let us consider a
contextual geometry $(O,C)$ and let us denote by $C^+$ the subset of positive
contexts, $C^-$ the subset of negative ones and by $\langle c\rangle$ the
expectation for an experiment corresponding to the context $c$. The following
inequality was established by A. Cabello~\cite{Cab10}:
 \begin{equation}
  \sum_{c\in C^+} \langle c\rangle-\sum_{c \in C^-} \langle c\rangle \leq b.
  \label{noncontextIneq}
 \end{equation}
Under the assumption of Quantum Mechanics, the upper bound $b$ is the number of
contexts of $(O,C)$, i.e. $b=l$. However this upper bound is lower for
contextual geometries under the hypothesis of a NCHV model. Indeed as shown
in~\cite{Cab10} this bound is $b=2s-l$ where $s$ is the maximum number of
constraints of the configuration that can be satisfied by an NCHV model. It
connects the notion of degree of contextuality with the upper bound $b$,
\begin{equation}\label{eq:context_inequality}
b=l-2d.\end{equation}

\section{Contextual configurations of the symplectic polar space}
\label{sec:symplectic-space}

We aim to automate the generation of contextual geometries and the detection of
their contextuality, with observables restricted to the elements of the
$n$-qubit Pauli group $P_n$, \textit{i.e.} the group of the $n$-fold tensor
products of Pauli matrices. We study them through their encoding as vectors from
a vector space over the two-elements field $\mathbb{F}_2$. This encoding does
not preserve all information, in particular we loose the commutation relations
and the signs of the contexts. At the geometrical level, the commutation
relation will be recovered by introducing a nondegenerate symplectic form
(\sec~\ref{sub:w_n}). The signs of the contexts will be determined as detailed
in \sec~\ref{sub:valuation}.

\subsection{The symplectic polar space \texorpdfstring{$W_n$}{Wn}}
\label{sub:w_n}

Let $V_n = \mathbb{F}_2^{2n}$ be the vector space of dimension $2n$ over the
two-elements field $\mathbb{F}_2$. Let $\bigotimes$ denote the (generalized)
tensor product. An element of the $n$-qubit Pauli group $P_n$ is an operator
$\mathcal{O}$ which can be factorized as $\mathcal{O} = s \bigotimes_{1 \leq i
\leq n} Z^{a_i} X^{b_i}$ with $a_i, b_i \in \mathbb{F}_2$ and a \emph{phase} $s
\in \{\pm 1, \pm i\}$. We denote by $C_n$ the center of $P_n$, \textit{i.e.}
$C_n = \{\pm Id, \pm i\,Id\}$. The surjective map $\pi : P_n \to V_n$ defined by
$\pi(s \bigotimes_{1 \leq i \leq n} Z^{a_i}X^{b_i})=(a_1, b_1, \ldots, a_n, b_n)$
factors through the isomorphism $\overline{\pi} : P_n/C_n \to V_n$ such that
$\overline{\pi} (\overline{\mathcal{O}}) = (a_1, b_1, \ldots, a_n, b_n)$, where
$\overline{\mathcal{O}}$ is the class of $\mathcal{O}$ in
$P_n/C_n$~\cite[\sec~2]{Hol21}.

$\overline\pi$ has several crucial aspects: its images are more elementary than
the original objects (binary vectors replace Hermitian matrices), and
$\overline\pi$ preserves some key properties about $P_n$. As defined,
$\overline\pi$ already transforms the matrix product into the sum in $V_n$. In
order to encode commutativity, we define the \emph{inner product} (also called
\emph{symplectic form}) on $V_n$ as $\innerproduct{x}{y} = xJy^\intercal,$
with
$$J = \begin{psmallmatrix} 
0&1           \\
1&0           \\
 & &\ddots    \\
 & &      &0&1\\
 & &      &1&0\\
\end{psmallmatrix}.$$ 
\noindent Then $\innerproduct{\overline\pi(\mathcal{\overline{O}})}{\overline\pi
(\mathcal{\overline{O'}})}=0$ iff the operators in $\mathcal{\overline{O}}$ and
$\mathcal{\overline{O'}}$ commute~\cite[\sec~2]{Hol21}.

Since the trivial $n$-qubit Pauli operator $Id$ does not correspond to a
measurement, we eliminate its class $\overline{Id}$ from $P_n/C_n$ and the
corresponding neutral element $\overline\pi(\overline{Id}) = (0,\ldots,0)$ from
$V_n$. This restriction of $\overline\pi$ is a bijection between the set of
classes of non trivial $n$-qubit Pauli observables and the \emph{projective
space} $PG(2n-1,2)$, whose points are nonzero vectors in $V_n$.

We can now define a counterpart to a quantum geometry in $PG(2n-1,2)$.
A \emph{quantum configuration} is a pair $(P,L)$ where $P$ is a finite set of
points of $PG(2n-1,2)$ and $L$ is a finite set of subsets of $P$, such that 
\begin{enumerate}[label=\textbf{S.\arabic*}]
  \item any two vectors $V$ and $W$ in the same element of $L$ commute, 
    \textit{i.e.} $\innerproduct{V}{W}=0$;
  \label{enum:nul-form}
  \item the sum of all vectors in each element of $L$ is $(0,\ldots,0)$.
  \label{enum:nul-sum}
\end{enumerate}

One can see that this definition of a quantum configuration corresponds through
$\overline\pi$ to that of a quantum geometry given in
\sec~\ref{sec:geometries}. Indeed Condition~\ref{enum:binary} is satisfied by
the elements of $P_n$, so the fact that we use only points from $PG(2n-1,2)$
satisfies it. Condition~\ref{enum:op-commut} is in correspondence with
Condition~\ref{enum:nul-form} and Condition~\ref{enum:context-prod} is in
correspondence with Condition~\ref{enum:nul-sum}.

A \emph{totally isotropic subspace} of $PG(2n-1,2)$ is a linear subspace $S$ of
$PG(2n-1,2)$ such that $\innerproduct{a}{b} = 0$ for any $a,b \in S$. Thus,
Condition~\ref{enum:nul-form} rewrites as ``all elements of $L$ are totally
isotropic subspaces''. The space of totally isotropic subspaces of $PG(2n-1,2)$
for $\innerproduct{\cdot}{\cdot}$ is called the \emph{symplectic polar space} of 
rank $n$ and order $2$ and is denoted by $W(2n-1,2)$ (abbreviated as $W_n$). The
\emph{points in} $W_n$ are the totally isotropic subspaces of dimension 0. In
all rigor they are the singletons $\{v\}$, for all points $v \in PG(2n-1,2)$,
but we identify them with their element $v$. They do not satisfy
Condition~\ref{enum:nul-sum}, whereas all other totally isotropic subspaces (of
positive dimension) satisfy it.

In this work, we only consider the pairs $(P,L)$ such that $P$ is a set of
points in $W_n$ and $L$ is a subset of $W_n$ composed of totally isotropic
subspaces with the same positive dimension. By construction, $(P,L)$ satisfies
Conditions~\ref{enum:nul-form} and~\ref{enum:nul-sum}, so it
is a quantum configuration.

\begin{example}
\label{ex:doily}
$W_2 = W(3,2)$, represented in \fig~\ref{fig:doily}, is the symplectic polar
space of rank $2$ and order $2$, corresponding to Pauli operators acting on two
qubits. It has 15 points (the subspaces of dimension 0) and 15 lines (the
subspaces of dimension 1), and no subspace of higher dimension. 
\forMinorRevision{This space, which is often represented by a pentagon-like shape
called \emph{the doily} (as in Figures~\ref{fig:doily} and
\ref{fig:doily-operators}), plays an important role in the quantum information
theory and the so-called black-hole-qubit correspondence. The reader interested
in learning more about these facts in a rather illustrative way can consult, for
example, Ref.~\cite{San19}, where s/he will also find further relevant
references on the topic.}

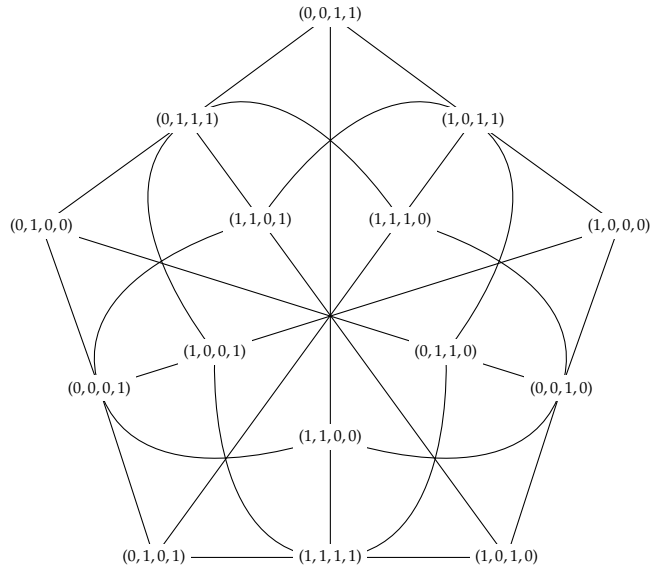
\begin{figure}[!ht]
\begin{center}
\begin{tikzpicture}[every plot/.style={smooth, tension=2},
  scale=4, 
  every node/.style={scale=0.6,fill=white}
]
\coordinate (1) at (0,1.0);
\coordinate (2) at (0,-0.80);
\coordinate (3) at (0,-0.40);
\coordinate (4) at (-0.95,0.30);
\coordinate (5) at (0.76,-0.24);
\coordinate (6) at (0.38,-0.12);
\coordinate (7) at (-0.58,-0.80);
\coordinate (8) at (0.47,0.65);
\coordinate (9) at (0.23,0.32);
\coordinate (10) at (0.58,-0.80);
\coordinate (11) at (-0.47,0.65);
\coordinate (12) at (-0.23,0.32);
\coordinate (13) at (0.95,0.30);
\coordinate (14) at (-0.76,-0.24);
\coordinate (15) at (-0.38,-0.12);

\draw (1) -- (11) -- (4);
\draw (4) -- (14) -- (7);
\draw (7) -- (2) -- (10);
\draw (10) -- (5) -- (13);
\draw (13) -- (8) -- (1);
\draw (7) -- (9) -- (8);
\draw (11) -- (12) -- (10);
\draw (13) -- (15) -- (14);
\draw (4) -- (6) -- (5);
\draw (1) -- (3) -- (2);
\draw plot coordinates{(15) (11) (9)};
\draw plot coordinates{(12) (8) (6)};
\draw plot coordinates{(9) (5) (3)};
\draw plot coordinates{(6) (2) (15)};
\draw plot coordinates{(3) (14) (12)};

\node at (1) {$(0,0,1,1)$};
\node at (2) {$(1,1,1,1)$};
\node at (3) {$(1,1,0,0)$};
\node at (4) {$(0,1,0,0)$};
\node at (5) {$(0,0,1,0)$};
\node at (6) {$(0,1,1,0)$};
\node at (7) {$(0,1,0,1)$};
\node at (8) {$(1,0,1,1)$};
\node at (9) {$(1,1,1,0)$};
\node at (10) {$(1,0,1,0)$};
\node at (11) {$(0,1,1,1)$};
\node at (12) {$(1,1,0,1)$};
\node at (13) {$(1,0,0,0)$};
\node at (14) {$(0,0,0,1)$};
\node at (15) {$(1,0,0,1)$};
\end{tikzpicture}
\end{center}
\caption{The doily, a point-line representation of $W_2=W(3,2)$, depicted in its 
standard rendering exhibiting a five-fold symmetry.
\modif{A point (i.e. a totally isotropic subspace of dimension 0) is represented
  by its coordinates in the ambient PG$(3,2)$, whereas a line is illustrated
  either as a straight segment or as a piece of an arc accommodating three
  distinct points. Note that the (modulo-two) sum of coordinates of the three
  points on any line is $(0,0,0,0)$.} \label{fig:doily}}
\end{figure}
\end{example}

\subsection{Context valuation, valuation vector and contextual configurations}
\label{sub:valuation}

Any quantum configuration $(P,L)$ in $W_n$ with $p = |P|$ points and $l = |L|$
(context) lines determines an incidence matrix $A \in \mathbb{F}_2^{l \times p}$
defined by $A_{i,j} = 1$ if the $i$-th element of $L$ contains the $j$-th point
in $P$. However, it provides no context/line valuation $e$, on which its
contextuality however depends. A context valuation can be derived as follows
from an interpretation of points in $W_n$ as Pauli operators, in other words
from a right inverse $\rho$ of the map $\pi$ ($\pi \circ \rho = id$). Among all
these inverses, we consider here the map $\rho : PG(2n-1,2) \to P_n$ defined by
$\rho\big((a_1,b_1,\dots,a_n, b_n)\big) = \bigotimes_i O_i$ with $O_i = Y$ if
$a_i=b_i=1$ and $O_i = Z^{a_i}X^ {b_i}$ otherwise. The corresponding context
valuation is the map $e_{\rho} : L \rightarrow \{-1,1\}$ such that $\prod_{p \in
l} \rho(p) = e_{\rho}(l)~Id$ for all $l \in L$.  It results from the commutation
relations on each context line that the values of $e_{\rho}(l)$ can only be $\pm
1$. The corresponding valuation vector $E_{\rho}$ is defined from $e_{\rho}$ as
in \sec~\ref{sec:degree}. Finally, a \emph{contextual configuration} is a
triple $(P,L,\rho)$ composed of a quantum configuration $(P,L)$ and a map $\rho$
from $PG(2n-1,2)$ to $P_n$, such that the linear system $A x = E_{\rho}$ has no
solution in $\mathbb{F}_2^{p}$. (In this definition, $A$ is the incidence matrix
of $(P,L)$.)

\begin{example}
\label{ex:doily-operators}
After replacing each point in $W_2$ by its image by $\rho$, \fig~\ref{fig:doily}
becomes \fig~\ref{fig:doily-operators}. The product of all observables on each
line is $I \otimes I$ (marked as a single black line) or $-I \otimes I$ (marked 
as a doubled red line). It determines the values $1$ and $-1$ of $e_{\rho}$. With
these  elements, it is well-known that the doily is a contextual configuration 
(see \textit{e.g.}~\cite{Cab10}).
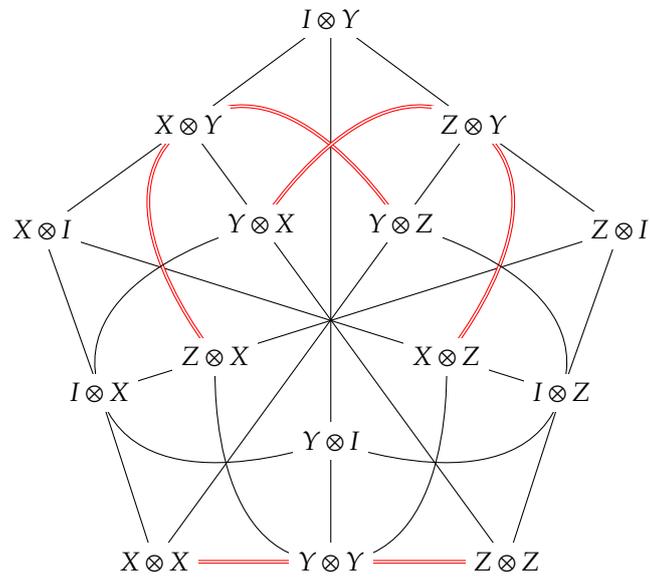
\begin{figure}[!ht]
\begin{center}
\begin{tikzpicture}[every plot/.style={smooth, tension=2},
  scale=4,
]
\coordinate (IY) at (0,1.0);
\coordinate (YY) at (0,-0.80);
\coordinate (YI) at (0,-0.40);
\coordinate (XI) at (-0.95,0.30);
\coordinate (IZ) at (0.76,-0.24);
\coordinate (XZ) at (0.38,-0.12);
\coordinate (XX) at (-0.58,-0.80);
\coordinate (ZY) at (0.47,0.65);
\coordinate (YZ) at (0.23,0.32);
\coordinate (ZZ) at (0.58,-0.80);
\coordinate (XY) at (-0.47,0.65);
\coordinate (YX) at (-0.23,0.32);
\coordinate (ZI) at (0.95,0.30);
\coordinate (IX) at (-0.76,-0.24);
\coordinate (ZX) at (-0.38,-0.12);

\draw (IY) -- (XY) -- (XI);
\draw (XI) -- (IX) -- (XX);
\draw[red,style=double] (XX) -- (YY) -- (ZZ);
\draw (ZZ) -- (IZ) -- (ZI);
\draw (ZI) -- (ZY) -- (IY);
\draw (XX) -- (YZ) -- (ZY);
\draw (XY) -- (YX) -- (ZZ);
\draw (ZI) -- (ZX) -- (IX);
\draw (XI) -- (XZ) -- (IZ);
\draw (IY) -- (YI) -- (YY);
\draw[red,style=double] plot coordinates{(ZX) (XY) (YZ)};
\draw[red,style=double] plot coordinates{(YX) (ZY) (XZ)};
\draw plot coordinates{(YZ) (IZ) (YI)};
\draw plot coordinates{(XZ) (YY) (ZX)};
\draw plot coordinates{(YI) (IX) (YX)};

\node at (IY) {$I \otimes Y$};
\node at (YY) {$Y \otimes Y$};
\node at (YI) {$Y \otimes I$};
\node at (XI) {$X \otimes I$};
\node at (IZ) {$I \otimes Z$};
\node at (XZ) {$X \otimes Z$};
\node at (XX) {$X \otimes X$};
\node at (ZY) {$Z \otimes Y$};
\node at (YZ) {$Y \otimes Z$};
\node at (ZZ) {$Z \otimes Z$};
\node at (XY) {$X \otimes Y$};
\node at (YX) {$Y \otimes X$};
\node at (ZI) {$Z \otimes I$};
\node at (IX) {$I \otimes X$};
\node at (ZX) {$Z \otimes X$};
\end{tikzpicture}
\end{center}
\caption{\modif{The doily of \fig~\ref{fig:doily} but with its points labeled by
  15 canonical representatives of the 15 equivalence classes of the two-qubit
  factor group $P_2/C_2$ (see also the dictionary in \sec~\ref{sub:subgeom} for
  more details).} \label{fig:doily-operators}}
\end{figure}
\end{example}

\section{Contextual subspaces of symplectic polar spaces}
\label{sec:contextual-subspaces}

\modif{This section presents an automatic method to generate contextuality
proofs. We first define programmable criteria to characterize candidate
geometries for contextuality. Then we describe our algorithm and code to
generate these geometries and detect their contextuality. Finally we sum up the
results obtained by execution of this code.}

\subsection{Selected subgeometries \add{and their group-theoretical counterparts}}
\label{sub:subgeom}

A \emph{line} of $W_n$ is a subspace of $W_n$ of (projective) dimension 1.
An element of $W_n$ of maximal dimension ($n-1$) is called a \emph{generator}. A
\emph{quadric} of $W_n$ is the set of points that annihilate a given quadratic
form. The most elementary quadratic form that we could define is $Q_0(x) =
x_1x_2 + \ldots + x_{2n-1}x_{2n}$ if $x =(x_1, x_2, \ldots, x_{2n})$. A
\emph{hyperbolic} quadratic form is a form $Q_p$ defined by $Q_p(x)=Q_0(x)+
\innerproduct{x}{p}$ where $p$ annihilates $Q_0$, whereas an \emph{elliptic}
form is such a form $Q_p$ where $p$ does not annihilate $Q_0$. Finally, a
\emph{hyperbolic} (resp. \emph{elliptic}) \emph{quadric} is a quadric
corresponding to the zero locus of a \emph{hyperbolic} (resp. \emph{elliptic})
quadratic form. The \emph{perpset} of the point $v\in W_n$ is the set of all
points $w$ isotropic to  $v$, \textit{i.e.} such that $\innerproduct{v}{w}=0$.

\begin{remark}
A \emph{(geometric) hyperplane} is a set of points in $W_n$ such that any line
of $W_n$ either has a single point of intersection with the hyperplane, or is
fully contained in it. The $10$ Mermin-Peres squares of $W_2$ are geometric
hyperplanes. It is also known that  the elliptic quadrics, the hyperbolic
quadrics and the perpsets are the only types of hyperplanes in
$W_n$~\cite{VL10}. In fact the $10$ Mermin-Peres squares are the $10$ hyperbolic
quadrics of $W_2$. This remark motivates our choice to study the contextuality
of all types of hyperplanes in $W_n$.
\end{remark}

Given these definitions, we consider the following families of geometries:
\begin{enumerate}[label=\textbf{G.\arabic*}]
  \item all points and all lines of $W_n$;
    \label{item:w-lines}
  \item all points and all generators of $W_n$;
    \label{item:w-blocks}
  \item all points and all lines in one hyperbolic quadric in $W_n$, for all 
    hyperbolic quadrics in $W_n$;
    \label{item:hyperbols}
  \item all points and all lines in one elliptic quadric in $W_n$, for all 
    elliptic quadrics in $W_n$;
    \label{item:ellips}
  \item all points and all lines in one perpset in $W_n$, for all perpsets of $W_n$.
    \label{item:perpsets}
\end{enumerate} 

\add{In what follows -- as already mentioned in \sec~\ref{sub:valuation} -- each 
equivalence class $\overline{\mathcal{O}}$ of the factor group $P_n/C_n$ will
be represented by a {\it single} element from that class, and namely by the 
canonical element with $s=1$ (see \sec~\ref{sub:w_n}). With this assumption in
mind, the correspondence between the structure of the factor group $P_n/C_n$
and that of the symplectic polar space $W_n$, described in general terms in 
\sec~\ref{sub:w_n}, acquires the following more explicit form:}

\begin{table}[h]
\resizebox{\textwidth}{!}{
\add{
\begin{tabular}{| c | c |}
  \hline
  $W_n$                             &  $P_n/C_n$                                                    \\
  \hline
  point                             & canonical element                                             \\
  \hline
  collinear points                  &  commuting canonical elements                                 \\
  \hline
  linear subspace of dimension $k$  & set of $2^{k+1}-1$ mutually commuting elements                \\
  ($k < n$)                         & whose product is $\pm I \otimes I \otimes ... \otimes I$      \\
  \hline                                  
  generator                         & maximal set of mutually commuting elements                    \\
  \hline
  perpset of a point                & set of elements commuting with a given element                \\
  \hline
  quadric associated with           & the set of symmetric/skew-symmetric elements                  \\
  a given point                     & commuting/anti-commuting with the corresponding $\mathcal{O}$ \\
  \hline                                      
  hyperbolic quadric                & $\mathcal{O}$ is symmetric                                    \\
  \hline
  elliptic quadric                  & $\mathcal{O}$ is skew-symmetric                               \\ 
  \hline
\end{tabular}
}
}
\end{table}

\add{\noindent Here we also add that a canonical element $\mathcal{O}$ is
symmetric or skew-symmetric if it contains an even or odd number of $Y$'s,
respectively. With the above-given group-geometric dictionary even the reader
who is unfamiliar with the terminology of finite geometry can fairly easily
grasp the concepts involved. By a way of illustration, let us take $W_2$, whose
15 canonical elements are depicted in \fig~\ref{fig:doily-operators}. One readily checks that, for
example, the hyperbolic quadric associated with the (symmetric) canonical
element $Y \otimes Y$ consists of the following nine canonical elements
$X \otimes X, Y \otimes Y, Z \otimes Z, X \otimes Z, Z \otimes X, X \otimes Y,
Y \otimes X, Y \otimes Z, Z \otimes Y$; here, the first five elements are
symmetric and each of them commutes with $Y \otimes Y$, whereas the remaining
four elements are skew-symmetric and none of them commutes with $Y \otimes Y$.
Similarly, the elliptic quadric associated, for example, with the
(skew-symmetric) canonical element $I \otimes Y$ is found to feature the
following five canonical elements $Y \otimes Y, X \otimes I, Z \otimes I,
Y \otimes X, Y \otimes Z$.}

\subsection{\add{Algorithms for contextuality proof generation}}
\label{sub:algo}

\add{This section presents our algorithms for the detection of contextual
configurations, and justifies their complexity.} \modif{They are implemented in
the language of the well-established tool Magma for theoretical mathematics.
Our Magma code is available on a public git whose page related to this
paper\footnote{\url{https://quantcert.github.io/Magma-contextuality}}
provides instructions on how to install and run it.} \add{In this paper the
functions we implemented in Magma are denoted in the \texttt{\textsc{Typewriter}}
font whereas the native Magma functions are denoted in the \textit{italic} font.}

\modif{The Magma language provides us with a function \textit{SymplecticSpace} to
build symplectic spaces.} \add{We simply wrap it in a custom function
\texttt{\textsc{QuantumSymplecticSpace}}$(n)$ that builds the points and the 
inner product $\innerproduct{\cdot}{\cdot}$ (introduced in \sec~\ref{sec:symplectic-space}) of
the symplectic space $W_n$ for $n$ qubits.} \add{
We also provide a function \texttt{\textsc{QuantumInc}} such that 
$\texttt{\textsc{QuantumInc}}(W_n)$ computes the collinearity relation between the
points of $W_n$ (generated by the function call 
\texttt{\textsc{QuantumSymplecticSpace}}$(n)$), and returns the incidence 
structure between these points and the cliques of the collinearity graph.}

\add{This function is the first building block of Algorithm~\ref{alg:subspaces},
that builds the subspaces of $W_n$ of any dimension $k$. Its main function is the
recursive function \texttt{\textsc{QuantumSubspaces}}, whose parameter is the 
dimension $k$ and that returns the set of all subspaces of $W_n$ of dimension
$k$. In the most elementary case, when $k = 0$, each subspace only consists of
one point of $W_n$ (Line~\ref{lst:line:basic}). Otherwise, each subspace of
positive dimension $k$ can be obtained by closing by summation an independent
set of $k+1$ collinear points. Line~\ref{lst:line:init} computes in $B$ the
maximal sets of mutually collinear points, that are the generators of $W_n$; it
is run outside the function body to compute it only once. Then, the two nested 
\textbf{for} loops build each set $blockSubset$ of $k+1$ collinear points. When
this set is independent (checked on Line~\ref{lst:line:check}), its closure for
the addition in $W_n$ is computed (on Line~\ref{lst:line:update}) by our generic
function \texttt{\textsc{Closure}}. Since the sum of a point with itself is $0$,
this null value is removed from the resulting set.}

\add{
\begin{algorithm}[!ht]
  \begin{algorithmic}[1]
  \State $B \gets Blocks(\texttt{\textsc{QuantumInc}}(W_n))$ \label{lst:line:init}
  \Function{\texttt{QuantumSubspaces}}{$n,k$}
  \If{$k = 0$}
    \State $subspaces \gets \{ \{point\} \mid point$ is a point of $W_n\}$ 
      \label{lst:line:basic}
  \Else
    \State $subspaces \gets \{\} $
    \State $previousSubspaces \gets \texttt{\textsc{QuantumSubspaces}}(k-1)$
    \ForEach{$block \in B$} \label{lst:line:block}
      \ForEach{subset $blockSubset$ of size $k+1$ of $block$} 
        \label{lst:line:blockSubset}
        \If{$blockSubset$ is not a subset of an element of $previousSubspaces$} 
          \label{lst:line:check}
          \State $subspaces \gets subspaces\cup\{\texttt{\textsc{Closure}}(blockSubset,+)
            \;\backslash\;\{0\}\}$ \label{lst:line:update}
        \EndIf
      \EndFor
    \EndFor
  \EndIf
  \State \Return $subspaces$
  \EndFunction
  \end{algorithmic}
  \caption{Generation algorithm for subspaces of dimension $k$ 
    of $W_n$. \label{alg:subspaces}}
\end{algorithm}
}

\add{This algorithm is very computationally demanding (iterating over the 
$\prod_{1 \leq i \leq n} (2^i+1)$ blocks, with several iterations nested in the 
outer one). In this paper, we only consider subspaces of dimension $1$ or $n-1$,
for which we propose the optimizations of Algorithm~\ref{alg:subspaces-special}.
For $k=1$ (lines of $W_n$, for the families of geometries~\ref{item:w-lines},
\ref{item:hyperbols}, \ref{item:ellips} and \ref{item:perpsets}), the function 
\texttt{\textsc{Lines}} returns a single geometry whose points are all points of
$W_n$ and whose contexts are the lines of $W_n$. Each line is composed of two
collinear points and their sum. For $k=n-1$ (generators of $W_n$, family of
geometries~\ref{item:w-blocks}), the function \texttt{\textsc{Generators}} 
returns a single geometry whose contexts are the blocks of the incidence
structure of $W_n$, computed by the native Magma function $Blocks$.}

\add{
\begin{algorithm}[!ht]
  \begin{algorithmic}[1]
  \Function{\texttt{Lines}}{$n$}
    \State $subspaces \gets \{\}$
     \ForEach{distinct points $p$ and $q$ in $W_n$}
      \If{$\innerproduct{p}{q} = 0$}
        \State $subspaces = subspaces \cup \{\{p,q,p+q\}\}$
      \EndIf
    \EndFor
  \State \Return $subspaces$
  \EndFunction
  \State
  \Function{\texttt{Generators}}{$n$}
    \State \Return $Blocks(\texttt{\textsc{QuantumInc}}(W_n))$
  \EndFunction
  \end{algorithmic}
  \caption{Effective generation algorithms for the subspaces of $W_n$
    of dimension $k = 1$ (lines) and $k= n-1$ (generators).
    \label{alg:subspaces-special}}
\end{algorithm}}

\add{Finally, the quadrics (families~\ref{item:hyperbols} and~\ref{item:ellips})
and perpsets (family~\ref{item:perpsets}) are computed by \forMinorRevision{
selecting lines of $W_n$ according to a criterion depending on one point $p$ of
$W_n$. For a quadric this criterion  is annihilation of a quadratic form:
$criterion(line,p)$ is $\forall \, point \in line, Q_p(point) = 0$. A line of a
perpset is incident with the point $p$ and its remaining two points are
orthogonal to $p$: $criterion(line,p)$ is $p \in line \text{ and } \forall\,
point \in line, \innerproduct{point}{p} = 0$. This selection} is formalized in
Algorithm~\ref{alg:quads-and-perps}.}

\add{
\begin{algorithm}[!ht]
  \begin{algorithmic}[1]
  \Function{\texttt{Quadrics/Perpsets}}{$n$}
    \State $lines \gets Lines(n)$
    \State $geometries \gets \{\}$
    \ForEach{point $p$ of $W_n$}
      \State $geometry \gets \{\}$
      \ForEach{$line \in lines$}
        \IIf{$criterion(line,p)$}
          {\;$geometry\gets geometry \bigcup \{line\}$}
        \EndIIf
      \EndFor
      \State $geometries \gets geometries \bigcup \{geometry\}$
    \EndFor
    \State \Return $geometries$
  \EndFunction
  \end{algorithmic}
  \caption{Quadrics and perpsets generation algorithms. 
  \label{alg:quads-and-perps}}
\end{algorithm}}

For all these geometries, we compute their line valuations, thanks to an
implementation of $\rho$. Then, contextuality is detected by \modif{building the
linear system described in \sec~\ref{sec:degree} and calling the Magma
function} \textit{IsConsistent} which determines whether a linear system has a
solution.

\modif{If the geometry is contextual, its} contextuality degree is evaluated by
a brute force algorithm which computes the Hamming distance between a given
vector $E\in \mathbb{F}_2^l$ and $\text{Im}(A)$ by evaluating this distance on
each vector of $\text{Im}(A)$ and selecting the smallest one. When the number of
contexts, i.e. the number of columns of $A$, increases this calculation becomes
quickly intractable. In~\cite{TLC22} an algorithm  with a quadratic speed-up --
based on techniques from error-correcting code theory -- is proposed to compute
the upper bound $b$ on non-contextuality inequalities (\ref{noncontextIneq}), 
which is equivalent to the computation of $d$, see 
\eq{\ref{eq:context_inequality}}.

\subsection{Complexity analysis}
\label{sub:complex}

\add{We present in this section a brief complexity analysis of our generation
algorithms. First of all, Magma documentation does not provide information about
the complexity of its functions \textit{SymplecticSpace} and
\textit{AllCliques} called by our code. We can ignore this complexity, since
these functions are called only once at initialization, and it has been
experimentally checked that they are not a bottleneck.}

For two qubits, the computation takes less than 0.1s, for three qubits the
computation takes around 5s. But for four qubits, the computation already takes
around 10min, and for five qubits, the computation takes around 24h.

These durations are consistent with the algorithmic complexity of the functions
computing the families of geometries, presented in \tab~\ref{tab:complexity}
and \modif{estimated} as follows.

\begin{table}[!ht]
\begin{center}
\begin{tabular}{|r|c|}
\hline
Geometries                        & Complexity                        \\
\hline
Lines (\ref{item:w-lines})        & $\bigO\left(2^{4n}\right)$        \\
Generators (\ref{item:w-blocks})  & $\bigO\left(2^{n(n+1)/2}\right)$  \\
Quadrics and Perpsets
(\ref{item:hyperbols} +
\ref{item:ellips} + 
\ref{item:perpsets})              & $\bigO\left(2^{6n}\right)$        \\
\hline
\end{tabular} 
\end{center} 
\caption{Algorithmic complexity for each geometry family.} 
\label{tab:complexity}
\end{table}

The symplectic space contains $2^{2n}-1$ points, so iterating over it is in
$\bigO(2^{2n})$. Each line contains three points, the third one being the sum of the
other two, so the complexity to generate all lines (for \ref{item:w-lines}) is
$\bigO\left(2^{2n} \times 2^{2n} \right) = \bigO\left(2^{4n}\right)$. This is consistent
with the number $(4^n-1) (4^{n-1}-1)/3)$ of lines in the symplectic space.

For the family \ref{item:w-blocks} of generators, we use the property that they
are the blocks of the incidence structure whose elements are the points of the
symplectic space and such that two points are incident if and only if they
\modif{are collinear}. The most expensive operation is the generation of the
$\prod_{1 \leq i \leq n} (2^i+1)$ blocks of this incidence structure, resulting 
in a complexity in $\bigO\left(2^{n(n+1)/2}\right)$.

\modif{The algorithm to compute the set of all quadrics iterates} over all the
points. For each point \modif{it generates} its quadratic form and
\modif{iterates} over all the points to find those who annihilate this quadratic
form. These points are the points of the quadric. The complexity of these two
operations is negligible compared to that of the next one: iteration over all
the lines of the symplectic space, a line being selected if it is a subset of
the points of the quadric. This yields a complexity in $\bigO\left(2^{2n}\times
2^{4n}\right)=\bigO\left(2^{6n}\right)$. The computation for the perpsets is very
similar.

\subsection{Results}
\label{sub:results}

\tab~\ref{tab:results} presents the contextuality results for a number of
qubits $2 \leq n \leq 5$. Each entry characterizes the contextual nature of the
corresponding geometries. The contextuality degree is provided when we succeeded
in computing it. Otherwise the letter {\bf C} indicates the contextual
configurations. The number of elements in the family is provided in parentheses.
Note that the elliptic quadrics for $n=2$, also known as ovoids, do not contain
any line. Therefore there are no contexts in this case and that is why we
indicate here "N/A". We also could have skipped the computation of the
generators for $n=2$, because their dimension is $n-1=1$ which means that
(\ref{item:w-lines})=(\ref{item:w-blocks}) for $n=2$. We kept it as it is
reassuring to see that we indeed obtain the same result for both families.

The cells in the {\bf bold font} provide contextuality results that
were not previously known. If the value is 0 then the family is non-contextual,
otherwise it is. There are three types of values obtained here: if we were able
to determine that a family was contextual without being able to compute the
contextuality degree, a "C" is written in the cell; if the degree was obtained
through computation, its value is simply given in the cell; at last, some cases
were not obtainable by computation, but can be derived from a closed formula
established in~\cite{Cab10}, in this case the value is recalled but the cell is
denoted by the {\it italic font}.

Some values were unobtainable due to the size of the systems. As mentioned
before the complexity of the naive algorithm to compute the distance between a
vector of $\mathbb{F}^l$ and the linear subspace $\text{Im}(A)\subset
\mathbb{F}^l$ increases exponentially with the rank of the matrix $A$. More
precisely the number of computations for an exhaustive search is
$\bigO(2^{\text{rank}(A)})$ and more sophisticated approaches only provide a
quadratic speed-up~\cite{TLC22}.

\begin{table}[!ht]
\def\nc{N}
\def\con{C}
\begin{center}
\begin{tabular}{|r|cccc|}
\hline
Geometries                        & $n=2$ & $n=3$     & $n=4$       & $n=5$       \\
\hline
Lines (\ref{item:w-lines})        & 3(1)  & \it90(1)  & \it1908(1)  & \it35400(1) \\
Generators (\ref{item:w-blocks})  & 3(1)  & \bf0(1)   & \bf0(1)     & \bf0(1)     \\
Hyperbolics (\ref{item:hyperbols})& 1(10) & \bf21(36) & \bf\con(136)& \bf\con(528)\\
Elliptics (\ref{item:ellips})     & N/A(6)& \bf9(28)  & \bf\con(120)& \bf\con(496)\\
Perpsets (\ref{item:perpsets})    & 0(15) & \bf0(63)  & \bf0(255)   & \bf0(1023)  \\
\hline
\end{tabular} 
\end{center}
\caption{Contextuality results for different values of the number of qubits $n$:
 For each family of geometries we provide either the degree of contextuality, or
 the letter {\bf C} to indicate that the configuration is contextual but the
 degree of contextuality is out of reach of our computational resources. We also
 provide in parentheses the number of occurrences in the given family of
 geometries. Results in the \textit{italic font} are deduced from general results
 regarding the upper bound $b$ while results in the \textbf{bold font} are obtained
 from our calculations.}
\label{tab:results}
\end{table}

The number of objects in each class was previously known, \cite{VL10}
gives a good overview of these numbers, which we recall and complete in
\tab~\ref{tab:cardinals} for convenience.

\begin{table}[!ht]
\begin{center}
\begin{tabular}{|r|ccc|}
\hline
Geometries $(P,L)$                & Cardinality of $P$  & Cardinality of $L$                                                          & Number of geometries\\ 
\hline
Lines (\ref{item:w-lines})        & $4^n-1$             & $(4^n-1)(4^{n-1}-1)/3$                                                      & 1                   \\
Generators (\ref{item:w-blocks})  & $4^n-1$             & $\prod_{1 \leq i \leq n}(2^i+1)$                                            & 1                   \\
Hyperbolics (\ref{item:hyperbols})& $(4^n+2^n)/2-1$     & $\left(\frac{4^n+2^n}{2}-1\right)\left(\frac{4^{n-1}+2^{n-1}}{2}-1\right)/3$& $(4^n+2^n)/2$       \\
Elliptics (\ref{item:ellips})     & $(4^n-2^n)/2-1$     & $\left(\frac{4^n-2^n}{2}-1\right)\left(\frac{4^{n-1}-2^{n-1}}{2}-1\right)/3$& $(4^n-2^n)/2$       \\
Perpsets (\ref{item:perpsets})    & $4^n/2-1$           & $4^{n-1}-1$                                                                 & $4^n-1$             \\
\hline
\end{tabular} 
\end{center} 
\caption{Cardinalities for each geometry family and their members.} 
\label{tab:cardinals}
\end{table}

\begin{remark} 
The contextuality of the configurations (\ref{item:w-lines}) has been
established in~\cite{Cab08}. We notice that the contextual nature of a given
configuration remains the same among all the geometries in the same family, and
for all sizes. For instance all hyperbolic quadrics are contextual. The only
exception is \ref{item:w-blocks} where this is not the case for $n=2$, but as
explained earlier, this comes from the fact that, in this case, generators are
in fact lines. From the geometric construction it is clear that for a fixed $n$,
the matrix $A$ in~\eq{\ref{eq:all-pc}} is the same (up to a change of basis)
for all geometries in the same family (it can also be seen from the fact that
the symplectic group $Sp(2n,2)$ acts transitively on the set of geometric
hyperplanes~\cite{VL10}). However the vector $E$ of the right-hand side 
of~\eq{\ref{eq:all-pc}} is not the same for all hyperbolic quadrics. To get a
better understanding of this, it will be interesting to understand for instance
how the symplectic group $Sp(2n,2)$ acts on the labeling of the contexts of
$W_n$.
\end{remark}

\begin{remark}
The contextual nature of the hyperbolic and elliptic quadrics for $n=3,4,5$ are
new results. In the $n=3$ case their computed degree of contextuality cannot be
explained by Theorem 15 of~\cite{TLC22}. It does not contradict this theorem,
which applies for graphs with an even number of contexts per point, but it
provides examples where the degree of contextuality is not the minimal number of
negative lines for all possible labelings of the configuration. Moreover
in~\cite{Hol21}, for the $n=3$ hyperbolic quadric, the bound
for~\eq{\ref{eq:context_inequality}} was estimated to be $51$ by counting the
minimal number of negative lines of the hyperbolic quadric, which is $27$. It
turns out that this bound $b^{NCHV}$ is in fact $b^{NCHV}=105-2\times 21=63$. It
does not change the conclusion of~\cite{Hol21}, as the experimental results of
the paper still violate the (now corrected) classical bound $b^{NCHV}$.
\end{remark}

\section{Contextuality degree of the quadrics of \texorpdfstring{$W_3$}{W3}}
\label{sec:quadric_geometry}

The new results provided in \tab~\ref{tab:results} concern the quadrics of
$W_n$ for $n=3,4,5$ in both hyperbolic and elliptic cases. Those quadrics are
contextual and their contextuality degree was obtained for $n=3$. In this
section we propose geometric evidence of this calculation by taking advantage of
geometric descriptions of those quadrics in $W_3$. \add{We also emphasize practical
applications of our work by describing some experimental characteristics for
testing quantum contextuality based on those quadrics. Those characteristics are
obtained from the degree of contextuality.}

\begin{figure}[!ht]
\begin{center}
\includegraphics[width=10cm]{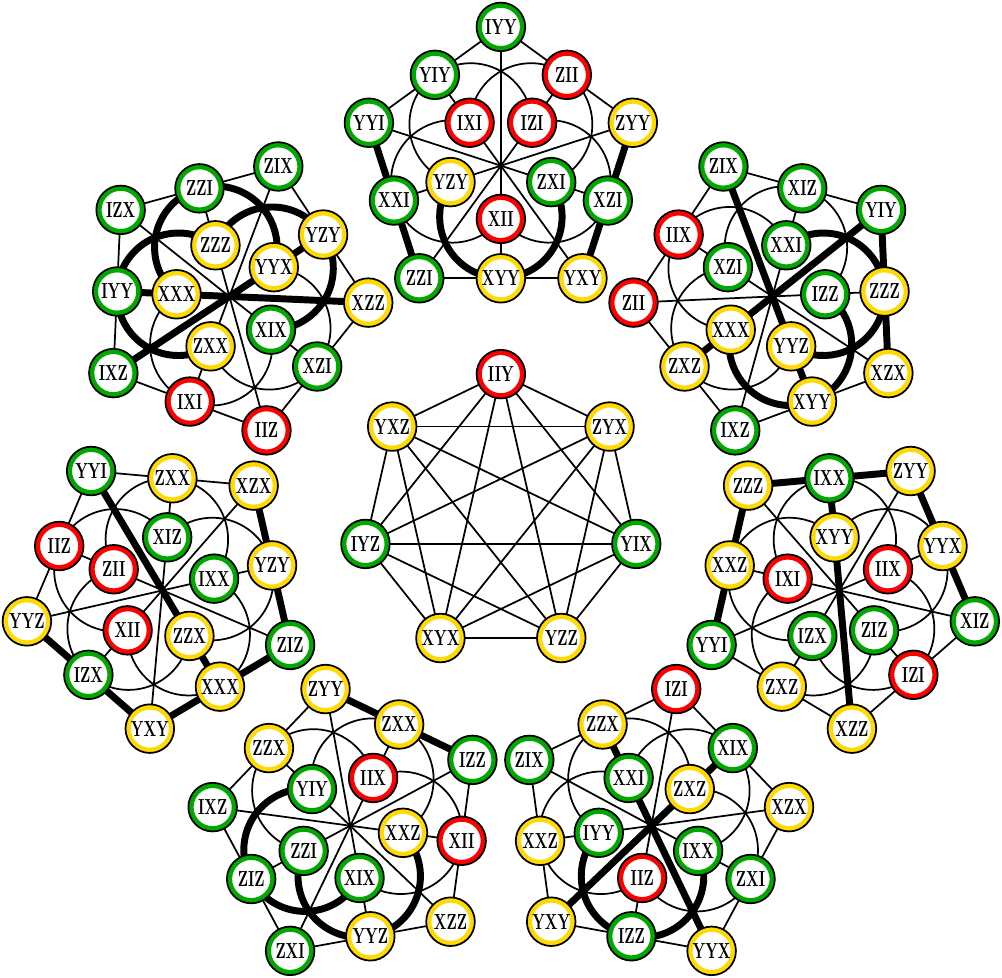}
\caption{\modif{An illustration of the structure of the three-qubit hyperbolic
  quadric $\mathcal{Q}_0$ in terms of seven doilies that pairwise share an
  elliptic quadric. There are shown all 105 lines (each belonging to a unique
  doily) and all 35 points (each occurring in three different doilies) of the
  quadric. Due to the lack of space, we use a shorthand notation $ABC$ instead of
  $A \otimes B \otimes C$ for the 63 canonical elements of $P_3/C_3$. The
  elements themselves are further distinguished by different colors according as
  they feature two $I$'s (red), one $I$ (green) aor no $I$ (yellow). The negative
  lines/contexts are boldfaced. The seven elements labeling the vertices of the
  central heptagon correspond to a particular Conwell heptad of points lying off
  $\mathcal{Q}_0$ (for more details, see~\cite{SdHG21}).} 
  \label{fig:heptad}}
\end{center}
\end{figure}

Recall that an example of a hyperbolic quadric is given by the zero locus of the
quadratic form $Q_0(x)$ with
\begin{equation}
 {Q}_0(x)=x_1x_2+\dots+x_{2n-1}x_{2n}.
\end{equation}
Let us denote by $\mathcal{Q}_0$ the quadric, i.e. the set of three-qubit Pauli
operators that annihilate ${Q}_0$. This quadric corresponds to the set of
symmetric three-qubit operators, i.e. three-qubit operators with an even number
of $Y$'s. \fig~\ref{fig:heptad} provides a line partition of $\mathcal{Q}_0$
into seven $W_2$, in which each of the $35$ operators appears in three different
$W_2$. The $105$ lines of $\mathcal{Q}_0$ correspond to the $7\times 15=105$
lines given by this line partition. This configuration has been previously 
introduced in~\cite[Section 5]{SdHG21} as a ‘Conwell’ heptad of doilies. It
provides an alternative description of the hyperbolic quadric $\mathcal{Q}_0$,
that we use here to compute the contextuality degree of $\mathcal{Q}_0$.

The degree of contextuality can be obtained by counting the minimal number of
constraints that cannot be satisfied by an NCHV model. It is always possible to
satisfy the constraints imposed by the $78$ positive lines of $\mathcal{Q}_0$
but the question of satisfying part of the constraints imposed by the $27$
negative lines is more subtle. The degree of contextuality of $W_2$ being $3$,
\fig~\ref{fig:heptad} implies that $7\times 3=21$ constraints cannot be
satisfied which corresponds exactly to the degree of contextuality of
$\mathcal{Q}_0$.

In the case of the elliptic quadric one can also give a nice justification of
the fact that their degree of contextuality is $9$ in $W_3$. Let us denote by
$\mathcal{Q}'$ such an elliptic quadric in $W_3$, i.e. the zero locus of an
elliptic quadratic form, see \sec~\ref{sec:contextual-subspaces}. It is known
that $\mathcal{Q}'$, as a point-line geometry, is a generalized quadrangle $GQ
(2,4)$, i.e. a point-line configuration of $3$ points per line and $5$ lines
per point which is triangle-free~\cite{SGL+10,LSVP09}. The configuration is made
of $27$ points and $45$ lines. As shown in~\cite{LSVP09}, out of those $45$ lines
one can build $36$ doilies such that every line shows up in $12$ different
doilies. But again one knows that in each doily there are always $3$ constraints
that cannot be satisfied because the degree of contextuality of $W_2$ is $3$.
Therefore the number of constraints of $GQ(2,4)$ that cannot be satisfied is
$36\times 3/12=9$, which is the degree computed in \sec~\ref{sec:quadric_geometry}.

To conclude this section, \modif{let us emphasize some practical information that
the knowledge of the contextual degree provides for experimental testing of
quantum contextuality. Beside the upper bound, $b$, for contextual inequalities
discussed in \sec~\ref{sec:degree}, let us discuss what} Cabello defines
in~\cite{Cab10} as a measure of robustness of a quantum violation of the
non-contextual inequality~\eq{\ref{eq:context_inequality}}. This notion of
{\em tolerated error per context} is expressed as
\begin{equation}
\varepsilon=\frac{N-b}{N}=\frac{2d}{N},
\end{equation}
where $N$ is the number of contexts of the configuration, $b$ is the upper bound
of~\eq{\ref{eq:context_inequality}} of an NCHV model and $d$ is the degree of
contextuality. \add{The tolerated error per context $\varepsilon$ measures the
errors that can occur, on each context, due to experimental imperfection, while
still violating the corresponding contextual inequality based on the
configuration.} \modif{Thanks to  the results obtained in this paper, we can
establish \tab~\ref{exptable}, that can be compared to \tab~I of~\cite{TLC22}.}
\add{One observes that the ratio $b/N$ and the tolerated error per context are as
good as the best contextual scenario proposed in \tab~I of~\cite{TLC22}.}
\begin{table}[!ht]
\begin{center}
 \begin{tabular}{|ccccc|}
  \hline
  Configuration       & Observables & Contexts  & $b/N$         & $\varepsilon$\\
  \hline
  Hyperbolic quadric  & $35$        & $105$     & $63/105=0.6$  & $0.4$        \\
  Elliptic quadric    & $27$        & $45$      & $27/45=0.6$   & $0.4$        \\
  \hline
 \end{tabular}
 \caption{\add{Characteristics for experimental implementations of contextual 
    inequalities based on $3$-qubit quadrics.}\label{exptable}}
\end{center}
\end{table}

From an experimental perspective, in order to obtain a violation 
of~\eq{\ref{eq:context_inequality}} we would like to have the smallest possible
ratio $\frac{b}{N}$ and the biggest value $\varepsilon$. In this respect both
hyperbolic and elliptic quadrics are as good as the Mermin pentagram which has a
smaller ratio $\frac{b}{N}$ and a bigger tolerated error $\varepsilon$ per
context than the new magic sets of~\cite{TLC22}.

\section{\add{Smallest split Cayley hexagons in \texorpdfstring{$W_3$}{W3}}}
\label{sub:cayley}

\add{
A particularly nice illustration of the power and physical relevance of our
contextuality algorithm is provided by another notable finite geometry living in
$W_3$, namely the split Cayley hexagon of order two, $\mathcal{H}$. Being a
generalized polygon like the doily itself (see, e.\,g., \cite{PSv01}), it is a
point-line incidence structure featuring 63 points and 63 lines, with three
points on a line and three lines through a point, whose group of automorphisms
is isomorphic to $G_2(2)$ of order $12\,096$. As it was first shown by Coolsaet
\cite{Coo10}, $\mathcal{H}$ can be embedded into $W_3$ in two fundamentally
different ways, called classical ($\mathcal{H}_{C}$; 120 distinct copies) and
skew ($\mathcal{H}_{S}$; 7560 copies). In a recent paper~\cite{HdS22}, three of
us employed the present algorithm to ascertain contextuality properties of each
of these three-qubit embedded split Cayley hexagons. The authors were quite
surprised to find out that although neither an $\mathcal{H}_{C}$ nor an
$\mathcal{H}_{S}$ is contextual, this is not the case with their
line-complements, $\overline{\mathcal{H}}$'s (note that $\mathcal{H}$ and
$\overline{\mathcal{H}}$ are identical as point-sets); indeed, it was found that
an $\overline{\mathcal{H}_{C}}$ is non-contextual, whereas any $\overline{
\mathcal{H}_{S}}$ is! So, this is the first stance of a quantum geometry ever
discovered where it is not so much the (properties of the) geometry {\it of its
own}, but rather the way how it {\it embeds} into the ambient multi-qubit
symplectic polar space that matters when it comes to quantum contextuality
issues. Moreover, given the fact that $\overline{\mathcal{H}_{S}}$'s exhibit a
considerable smaller degree of symmetry than $\overline{\mathcal{H}_{C}}$'s, this
finding also seems to indicate that in our future quest(s) for other examples of
contextual configurations one should pay a particular attention to those finite
geometries that are not so symmetry-pronounced.}

\section{Conclusion}
\label{sec:conclusion}

In this paper we performed a systematic computer-aided study of observable-based
proofs of contextuality with large numbers of contexts and observables as
subgeometries of the symplectic polar space $W_n$. We generate in particular
(in~\sec~\ref{sec:contextual-subspaces}) new proofs based on hyperbolic and
elliptic quadrics for $n=3,4,5$. \add{Moreover, we quantified contextuality
through the notion of contextuality degree and we computed this degree for these
new proofs, in the $n=3$ case.}

A natural generalization would be to prove that quadrics are always point-line
configurations that provide  observable-based proofs of the Kochen-Specker
Theorem for any $n$. The fact that for $n=3$ we were able to compute the degree 
of contextuality for both $\mathcal{Q}_0$, hyperbolic quadric, and 
$\mathcal{Q}'$, elliptic quadric, by geometric partition of the lines in terms of
doilies can be considered as a concrete motivation for studying how doilies cover
$W_n$ and its subspaces. In this respect our work on the taxonomy of the
symplectic polar spaces by means of $W_2$~\cite{SdHG21} can be used to estimate
the degree of contextuality for subgeometries of $W_n$. Those questions will be
addressed in a future work.

\section*{Acknowledgments}
\label{sec:acknowledgments}

This project is supported by the French Investissements d'Avenir program, project
ISITE-BFC (contract ANR-15-IDEX-03), and by the EIPHI Graduate School (contract
ANR-17-EURE-0002). The computations have been performed on the supercomputer
facilities of the Mésocentre de calcul de Franche-Comté. This work also received
a partial support from the Slovak VEGA grant agency, Project 2/0004/20. We also
thank our friend Zsolt Szabó for the help in preparing \fig~\ref{fig:heptad}.

\bibliographystyle{alpha}
\newcommand{\etalchar}[1]{$^{#1}$}

\end{document}